\documentclass[useAMS,referee,usenatbib]{biom}
\usepackage{graphicx}
\usepackage{color}
\usepackage{graphicx}
\usepackage{natbib}

\usepackage{amsmath}
\usepackage{hyperref}

\usepackage{url}
\makeatletter
\g@addto@macro{\UrlBreaks}{\UrlOrds}
\makeatother
\def\bSig\mathbf{\Sigma}

\newcommand{\R}{\mathbf{R}}
\newcommand{\indep}{\rotatebox[origin=c]{90}{$\models$}}
\newcommand{\NDE}{\textnormal{NDE}}
\newcommand{\NIE}{\textnormal{NIE}}

\newcommand{\TE}{\textnormal{TE}}

\newcommand{\bmm}{\mbox{\boldmath $m$}}
\newcommand{\bmM}{\mbox{\boldmath $M$}}
\newcommand{\bmW}{\mbox{\boldmath $W$}}
\newcommand{\bmw}{\mbox{\boldmath $w$}}

\newcommand{\bmX}{\mbox{\boldmath $X$}}
\newcommand{\bmx}{\mbox{\boldmath $x$}}

\newcommand{\bmalpha}{\mbox{\boldmath $\alpha$}}
\newcommand{\bmbeta}{\mbox{\boldmath $\beta$}}
\newcommand{\bmeta}{\mbox{\boldmath $\eta$}}

\title[Joint Model for Causal Mediation Analysis]{Joint Model for Mediation Analysis with Causally Related Longitudinal and Recurrent Event Mediators for Survival Outcome}

\author{Fang Niu$^{1}$,
Cheng Zheng$^{1,*}$\email{cheng.zheng@unmc.edu}, and
Lei Liu$^{2}$ \\
$^{1}$Department of Biostatistics, University of Nebraska Medical Center, Omaha, Nebraska, U.S.A. \\
$^{2}$Division of Biostatistics, Washington University in St. Louis, St. Louis, Missouri, U.S.A.\\
}

\begin{document}

%  This will produce the submission and review information that appears
%  right after the reference section.  Of course, it will be unknown when
%  you submit your paper, so you can either leave this out or put in
%  sample dates (these will have no effect on the fate of your paper in the
%  review process!)

\date{}

%  These options will count the number of pages and provide volume
%  and date information in the upper left hand corner of the top of the
%  first page as in published papers.  The \pagerange command will only
%  work if you place the command \label{firstpage} near the beginning
%  of the document and \label{lastpage} at the end of the document, as we
%  have done in this template.

%  Again, putting a volume number and date is for your own amusement and
%  has no bearing on what actually happens to your paper!

\pagerange{}
\volume{}
\pubyear{}
\artmonth{}

%  The \doi command is where the DOI for your paper would be placed should it
%  be published.  Again, if you make one up and stick it here, it means
%  nothing!

%\doi{10.1111/j.1541-0420.2005.00454.x}

%  This label and the label ``lastpage'' are used by the \pagerange
%  command above to give the page range for the article.  You may have
%  to process the document twice to get this to match up with what you
%  expect.  When using the referee option, this will not count the pages
%  with tables and figures.

\label{firstpage}

%  put the summary for your paper here

\begin{abstract}
Recurrent events and repeated measures are commonly encountered in clinical longitudinal studies, often holding strong associations with patient outcomes. Although joint models for repeated measures, recurrent events, and a terminal event have been developed to account for their correlation, limited methodologies exist to examine causal mediation mechanisms involving multiple types of mediators, especially when mediators are causally related. This study addresses this gap by proposing a novel causal mediation analysis framework to quantify natural direct and indirect effects when both recurrent events and repeated measures act as mediators with causal dependencies. We extend joint modeling approaches by incorporating shared random effects (frailties) structures, relaxing the commonly used ``sequential ignorability" assumption, and accounting for unmeasured time-independent confounders through shared random effects. We apply our method to the Terry Beirn Community Programs for Clinical Research on AIDS (CPCRA) study and demonstrate that both recurrent opportunistic infections (OIs) and repeated CD4 measurements mediate the effects of prior AIDS-defining conditions on survival outcomes. Additionally, the shared random effects between repeated CD4 and survival models highlight the presence of unmeasured confounding between CD4 counts and mortality. Simulation studies demonstrate the robustness and finite sample performance of our estimators for natural direct and indirect effects. The proposed methodology enables a more comprehensive investigation of causal pathways in longitudinal studies with multiple mediators, providing insights into treatment mechanisms and informing clinical decision-making.
\end{abstract}

%  Please place your key words in alphabetical order, separated
%  by semicolons, with the first letter of the first word capitalized,
%  and a period at the end of the list.
%

\begin{keywords}
causal inference; joint modeling; multiple mediation analysis; recurrent event; repeated measurement; survival data
\end{keywords}

%  As usual, the \maketitle command creates the title and author/affiliations
%  display

\maketitle

%  If you are using the referee option, a new page, numbered page 1, will
%  start after the summary and keywords.  The page numbers thus count the
%  number of pages of your manuscript in the preferred submission style.
%  Remember, ``Normally, regular papers exceeding 25 pages and Reader Reaction
%  papers exceeding 12 pages in (the preferred style) will be returned to
%  the authors without review. The page limit includes acknowledgements,
%  references, and appendices, but not tables and figures. The page count does
%  not include the title page and abstract. A maximum of six (6) tables or
%  figures combined is often required.''

%  You may now place the substance of your manuscript here.  Please use
%  the \section, \subsection, etc commands as described in the user guide.
%  Please use \label and \ref commands to cross-reference sections, equations,
%  tables, figures, etc.
%
%  Please DO NOT attempt to reformat the style of equation numbering!
%  For that matter, please do not attempt to redefine anything!

\section{Introduction}
\label{s:intro}

Longitudinal data with recurrent events and repeated measures are prevalent across various fields and are increasingly common in clinical trials. Collecting repeated measurements of key variables and tracking recurring events provide a comprehensive view of patient conditions over time, offering valuable insights into their relationship with clinical outcomes. Examples of recurrent events include tumor relapses in breast cancer, recurrent hospitalizations in heart failure, and recurring opportunity infections (OIs) in HIV. Repeated measures involve the longitudinal acquisition of data from the same subjects, typically focusing on crucial biomarkers for monitoring disease progression. For instance, breast cancer patients routinely undergo blood tests to assess CA 15-3 or CA 27.29 levels, individuals with heart failure monitor blood pressure and cholesterol levels \citep{Deng2022, Lahoz2022}, and HIV patients receive  regular assessments of CD4 T-cell lymphocyte count (CD4 count). As technology advances, the concurrent collection of both types of data is increasingly common \citep{May2016}. These variables often hold strong associations with patient outcomes, e.g., mortality and disease diagnosis \citep{Djawe2015, Rack2010}. Understanding the relationships among recurrent events, repeated measures, and terminal events has stimulated significant interest in identifying prognostic biomarkers, developing novel therapeutic strategies, and establishing relevant clinical endpoints. However, suitable methods to rigorously investigate the roles played by recurrent events and repeated measures within the causal pathways linking exposure to time-to-event outcomes remain underdeveloped.

In the context of HIV-infected patients, the repeated occurrence of OIs poses a substantial adverse impact on health and survival, particularly in those diagnosed with AIDS (Acquired Immunodeficiency Syndrome). Furthermore, tracking CD4 counts before and after combination antiretroviral therapy (cART) is crucial for patient management, as it is strongly associated with OIs and mortality \citep{Hoffmann1999}. In Terry Beirn Community Programs for Clinical Research on AIDS (CPCRA) \citep{Abrams1994, Neaton1994}, 467 HIV-infected patients with CD4 $<$ 300 $cells/mm^3$ received treatments didanosine (ddI) or zalcitabine (ddC) for 1 to 21 months (median: 13 months). CD4 cell counts were measured at baseline and every two months during follow-up. There are 188 patients who died, and 363 cases of confirmed or probable opportunistic diseases documented. Notably, each patient experienced a variable number of OIs, ranging from zero to five. Previous studies demonstrated a strong positive correlation between recurrent OIs and risk of mortality\citep{Liu2008b, Niu2023}. Additionally, higher repeated CD4 cell counts were found associated with a decreased risk of recurrent OIs \citep{liu2009}. Concurrently, baseline CD4 counts were associated with both survival and OI risk, highlighting the interplay among OIs, CD4 counts, and patient mortality.
In the current study, our primary objective is to develop methodologies for causal mediation analysis to quantitatively assess the extent to which treatment effectiveness or effect of baseline conditions (e.g., prior AIDS-defining conditions) on overall survival is mediated by OI occurrences or repeated CD4 counts measurements. We also aim to elucidate the role of CD4 counts in the pathway through which OIs mediated treatment effects on patient outcomes.

\begin{figure}
\includegraphics[scale=0.8]{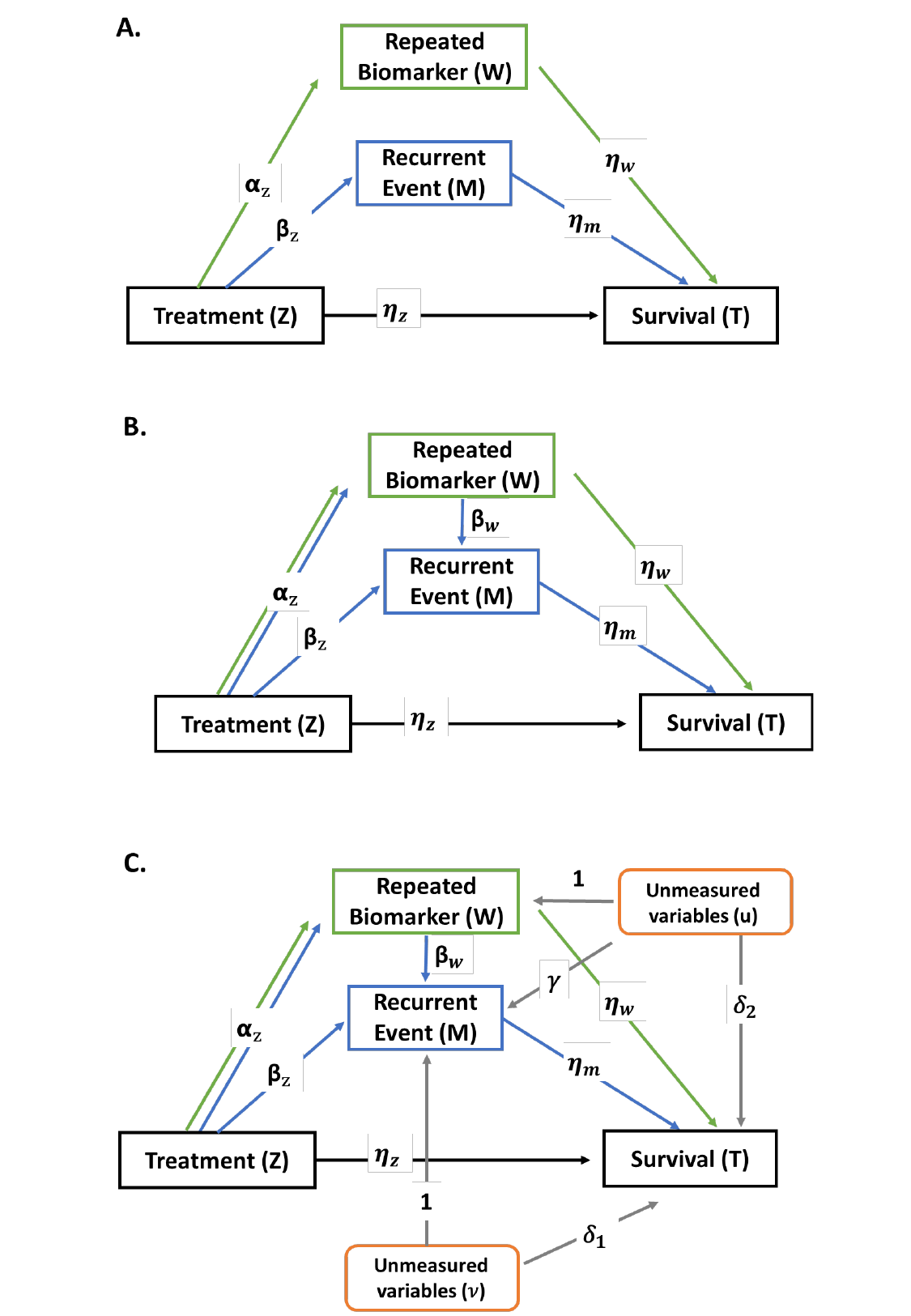}
\caption{Directed acyclic graph for commonly encountered causal mechanisms among the exposure (treatment $Z$), the recurrent event (mediator $M$), repeated biomarker (alternative mediator $W$), and the terminal event (outcome $T$). The direct effect of Treatment $Z$ on Survival $T$ is represented in the black arrow. The indirect effect from Treatment $Z$ to Survival $T$ mediated through the recurrent event $M$ and through the sequential mediation of repeated biomarker $W$ and recurrent event $M$ is represented in the blue arrow. The indirect effect of Treatment $Z$ on Survival $T$ mediated solely through the repeated biomarker $W$ is represented in the green arrow. }
\label{fig:1}
\end{figure}

Existing methodologies have limited ability to handle causal mechanisms with multiple mediators of interest. A common approach is analyzing each mediator separately, which assumes no relationships among mediators. However, mediators frequently influence outcomes and are interrelated, challenging the assumption of independence. Studies evaluating recurrent event mediators and repeated measuring mediators separately \citep{liu2009,Paulon2020,proustlima2009,vansteelandt2019} highlight the need for methodologies that jointly analyze correlated mediators while addressing unmeasured confounding.
Imai developed a sensitivity analysis for examining different alternative assumptions when multiple, causally related mediators exist\citep{Imai2013}. 
We extended the scenarios they mentioned and proposed a framework to analyze multiple mediators. 
Figure \ref{fig:1} illustrates causal diagrams with multiple mediators. The treatment variable $Z$ could affect the survival outcome $T$ via three different pathways: 1) through the direct pathway: $Z$ $\rightarrow$ $T$; 2) through the main mediator of interest $M$ (may or may not through other mediators): $Z$ $\rightarrow$ $M$ $\rightarrow$ $T$; 3) through only the alternative mediator $W$ but not through $M$: $Z$ $\rightarrow$ $W$ $\rightarrow$ $T$. In Figure 1A, mediators $M$ and $W$ are assumed to have no causal relationship with each other, while Figure 1B allows the mediator $M$ to be causally affected by the alternative mediator $W$. We adopt the latter scenario as it better represents our study.

Moreover, the relationships between mediators and the survival outcome $T$ may be influenced by confounding variables. In Figure 1B, there is no unmeasured confounding among the mediators, terminal events, or between the mediators themselves. In Figure 1C, the mediators serve as causal factors for the terminal event, but their association is partially confounded by unmeasured variables. Additionally, the relationship between the mediator $M$ and the alternative mediator $W$ is also confounded. These scenarios highlight the importance of addressing unmeasured confounding.

To handle unmeasured confounding in mediation analysis, prior studies applied a relaxed version of the ``sequential ignorability" (SI) assumption. SI assumption implies (i) conditioning on observed covariates, there is no confounding between the exposure and the mediator process; and (ii) conditioning on observed covariates and the exposure, there is no confounding between the mediator process and the outcome of interest. The relaxation assumption of SI allows unmeasured confounding between mediator and outcome by incorporating these unmeasured confounding variables into models. Recent methods showed that using factor analysis or shared random effects could adjust for some of the unmeasured confounding and thus relax the SI assumption \citep{liu2018, wang2019, mckennan2019, zheng2021}. 
This study applies a novel joint random effects (frailty) model of repeated-measuring mediators, recurrent events, and a survival outcome, to quantify the natural direct effect ($\NDE$) and natural indirect effect for the mediators $M$ ($\NIE_M$) and the alternative mediator $W$ ($\NIE_W$) while controlling certain unmeasured confounding under the relaxed SI assumption.

In the causal mediation framework, indirect effects represent the impact mediated through mediators, while direct effects capture impacts bypassing mediators. In our study, $\NDE$ captures the direct effects of the exposure $Z$ on the survival outcome $T$, while the indirect effect $\NIE_M$ accounts for the influence of the exposure on the survival outcome mediated through the recurrent event process $M$ (both directly through $M$ or through the pathway from the alternative mediator $W$ to $M$). $\NIE_W$ represents the indirect effect of the exposure on the survival outcome mediated only via the repeated measure $W$. Several methodologies for defining the causal mediation effect when the longitudinal biomarker or recurrent events mediators exist under semi-competing risk settings or joint modeling \citep{huang2021, gao2023, Stensrud2022,didelez2019, zheng2021, zheng2017}. However, research addressing causality in jointly analyzing multiple mediators and terminal event data remains limited.

This paper introduces estimators for $\NDE$, $\NIE_M$, and $\NIE_W$ when the recurrent event mediator, repeated measuring mediator, and the terminal event outcome are jointly modeled under the relaxed SI assumption that incorporates shared random effects between mediators and outcomes. Specifically, we focus on the shared frailty model that extends the joint model presented in precious studies \citep{liu2004,Niu2023}. As closed-form solutions are unavailable, we utilize Monte Carlo numerical integration for random effects. Finite sample performance and sensitivity to model misspecification are assessed through simulations for $\NDE$, $\NIE_M$, and $\NIE_W$ estimators. Our methods are applied to the CPCRA study, separately estimating the effects of two exposures: (i) ddI/ddC, (ii) PADC (prior AIDS-defining conditions), on patients' survival. We evaluate the mediation effects of recurring OIs and repeated measuring of CD4 cell counts in the pathway of the exposures separately on the survival outcome.

The rest of the paper is organized as follows. Section 2 defines notations and presents the formula for calculating the $\NDE$, $\NIE_M$, and $\NIE_W$ followed by a discussion of the numerical computation approach. Section 3 presents simulation studies to explore the finite sample performance of our proposed method. Section 4 applies our method to the CPCRA study. Section 5 summarizes the findings and proposes extensions.

\section{Methods}
\label{sec2}
We adopt the potential outcomes framework to define and estimate causal effects in our analysis. Within this framework, the extended version of the Stable Unit Treatment Value Assumption (SUTVA) is assumed for our mediation analysis. SUTVA asserts that the treatment assigned to one individual does not affect the outcome of another individual, and the potential outcomes for each individual are the same under the same treatment level, regardless of how the treatment is administered. SUTVA has been extended to include the mediators in mediation analysis. Specifically, this extended assumption indicates that the potential outcome for each individual is uniquely determined under a specified combination of treatment and mediator levels, independent of how the treatment and mediator are distributed among other individuals. Under the potential outcomes framework and SUTVA assumption, we denote the treatment assignment or exposure level of individual $i$ as $Z_i$, the repeatedly measured potential biomarker mediator at time $s$ under potential treatment $z$ as $W_{i}(z, s)$, and the corresponding biomarker mediator process $\underline{W}_{i} (z)={W_i(z, s): s \in [0, \tau]}$. %We also define the potential counting process of recurrent events (as mediator M) for individual $i$ at time $t$ if assigned treatment $z$ and time-varying mediator $\underline{w}$ as $M_{i}^{z}(t,\underline{w})$, where $t$ is in the interval $[0,\tau]$. 
We use $n_{i}(z,\underline{w})$ to represent the number of potential recurrent events that occur within $[0,\tau]$ under treatment $z$ and repeatedly measured mediator $\underline{w}$. If the potential recurrent event times for individual $i$ are $0<R_{i1}(z,\underline{w})<R_{i2}(z,\underline{w})<\cdots<R_{in_{i}(z,\underline{w})}(z,\underline{w})\leq\tau$, we have $M_i(z,t,\underline{w})=\sum_{j=1}^{n_{i}({z,\underline{w}})} I(R_{ij}(z,\underline{w})\leq t)$, where $j=1$ to $n_{i}(z,\underline{w})$ represents the potential counting process of recurrent events (as mediator $\bmM$) at time $t$ if assigned treatment $z$ and repeated measuring mediator $\underline{w}$. We define $T_i(z,\underline{m}, \underline{w})$ as the potential time to the terminal event with treatment $z$, recurrent event mediator counting process $\underline{m}=\{m(t,\underline{w}),t \in [0,\tau]\}$, and the repeatedly measured potential mediator $\underline{w}=\{w(s),s \in [0,\tau]\}$. We denote $C_i(z,\underline{m},\underline{w})$ as the corresponding potential censoring time. For our data examples, we cannot measure the mediator $\bmW$ continuously, and $W_i(s)$ are only observed at $s_{i1}, \cdots, s_{il_i}$. 

In this study, we assume that recurrent events $M$ and repeatedly measured mediator $W$ cannot happen after death. Thus, if the treatment has a direct effect on the potential survival time 
$T_i(z,\underline{M}_i(z',\underline{W}_i(z')), \underline{W}_i(z''))$ and the potential survival time under $z$ ($T_i(z,\underline{M}_i(z, \underline{W}_i(z)), \underline{W}_i(z)$) is shorter than the potential survival time under $z'$ 
($T_i(z',\underline{M}_i(z, \underline{W}_i(z)), \underline{W}_i(z)$), then the mediator processes $\underline{M}_i(z,\underline{W}_i(z))$ and $\underline{W}_i(z)$ will end earlier and become incomplete.  To address this issue, we assume the treatments for survival, recurrent event processes, and repeated measuring potential mediators can be hypothetically separated and won't affect each other. 
Thus, the recurrent events process and the time-varying potential mediators' process under certain treatment when the subject is alive can be defined, and $T_i(z,\underline{M}_i(z',\underline{W}_i(z')), \underline{W}_i(z''))$ represents the survival event under treatment $z$, recurrent events counting process under treatment $z'$ and repeated measuring mediator under treatment $z''$. 

In this study, we focus on the assessment of mediation involving repeatedly measured and counting process mediators simultaneously. To effectively handle these two types of mediators, we need to understand the potential relationship between these two mediators. Under our proposed models, we assume the mediator of $M$ is allowed to depend on a set of $W$. This means that the potential values of $M$ are a function of $W$.

The consistency assumption requires no measurement error in the mediator process. This assumption is reasonable when the recurrent events and repeatedly measured mediators are recorded timely and accurately. This assumption links the potential outcomes to the observed outcomes. We have potentially observable recurrent mediators counting process $M_i(t,\underline{W}_i)=M_{i}(t,Z_{i},\underline{W}_i(Z_i))$, repeated measuring mediators $W_i(s)=W_i(s, Z_i)$, time to the terminal event $T_i=T_i(Z_i,\underline{M}_i,\underline{W}_i)$, and censoring time $C_i=C_i(Z_i,\underline{M}_i, \underline{W}_i)$, where $\underline{M}_i= \{M_{i}(t,Z_{i},\underline{W}_i(Z_i)),t \in [0,\tau]\}$ and $\underline{W}_i= \{W_{i}(s,Z_{i}),s \in [0,\tau]\}$.  Due to censoring, we can only observe the follow-up time $T_i^{\ast}=T_i\wedge C_i$ and the terminal event indicator $\Delta_i=I(T_i\leq C_i)$. We denote the observed baseline covariates for individual $i$ as $\bmX_i \in \R^p$.

Within this framework, we define the direct effect $\NDE$ and indirect effects $\NIE_M$ and $\NIE_W$ based on their relationships described in Figure \ref{fig:1}. As shown in Figure \ref{fig:1} C, $\NDE$ represents the direct effect of Treatment $Z$ on Survival $T$ (black arrow). The indirect effect $\NIE_M$ captures two effect pathways, one is from Treatment $Z$ to Survival $T$ mediated through the recurrent event $M$; the other one is from Treatment $Z$ to Survival $T$ mediated first through the repeated biomarker $W$ and then recurrent event $M$ (blue arrow). Similarly, $\NIE_W$ denotes the direct effect of Treatment $Z$ on Survival $T$ mediated solely through the repeated biomarker $W$ (green arrow). These direct and indirect effects can be estimated on various scales, such as the survival function, cumulative hazard function, or hazard function. 
In this paper, we focus on the survival function scale and define these effects as:
\begin{eqnarray*}
\NDE(t,z,z',z'')&=&Pr(T(z',\underline{M}(z, \underline{W}(z)),\underline{W}(z'')) > t) -Pr(T(z,\underline{M}(z,\underline{W}(z)),\underline{W}(z'')) > t),\\
\NIE_M(t,z,z',z'')&=&Pr(T(z',\underline{M}(z',\underline{W}(z')),\underline{W}(z'')) > t) -Pr(T(z',\underline{M}(z,\underline{W}(z)),\underline{W}(z'')) > t),\\
\NIE_W(t,z,z',z'')&=&Pr(T(z',\underline{M}(z'',\underline{W}(z'')),\underline{W}(z')) > t) -Pr(T(z',\underline{M}(z'',\underline{W}(z'')),\underline{W}(z)) > t),
\end{eqnarray*}
where $z=0,1$, $z'=0,1$ and $z''=0,1$. We can see that the total effect can be decomposed into the sum of these three effects as below:
$$
\TE(t,z,z')=\NDE(t,z,z',z) + \NIE_M(t,z,z',z') + \NIE_W(t,z,z',z).
$$

Here, $\NDE(t,z,z',z'')$ is the direct effect of the exposure on the survival outcome contrasting level $z$ to $z'$, while fixing the mediator $\bmM$ at its potential level under exposure $z$ (i.e., $\underline{M}(z,\underline{W}(z))$) and the mediator $\bmW$ at its potential level under the exposure $z''$ (i.e., $\underline{W}(z'')$); $\NIE_M(t,z,z',z'')$ is the indirect effect of the exposure on the survival outcome via recurrent events mediator $\bmM$ contrasting level $z$ to $z'$, while controlling the potential exposure's direct impact on the survival outcome T under level $z'$ and the mediator $\bmW$ under its potential level under exposure $z''$ (i.e., $\underline{W}(z'')$); $\NIE_W(t,z,z',z'')$ is the indirect effect of the exposure on the survival outcome via repeatedly measured mediator $\bmW$ contrasting level $z$ to $z'$, while controlling the potential exposure's direct impact on the survival outcome T under the potential exposure $z'$ and the mediator $\bmM$ at its potential level under the exposure $z''$ (i.e., $\underline{M}(z'',\underline{W}(z''))$). Details that apply these functions to compute $\NDE$, $\NIE_M$, and $\NIE_W$ can be found in Web Appendix B.

%{\color{red}[This equation seems not hold since we have the direct effect is the NDE defined above is the effect that are not through either $M$ and $W$ while $NIE_M$ are effects through $M$, so the effect that through $W$ but not through $M$ are not counted in either term and thus the sum will not be the TE which defined as:}

%{\color{blue}[It seems that we can have 

%if we either modify the definition of $\NIE_M(t,z,z')$ as $$\NIE_M(t,z,z')=Pr(T(z',\underline{M}^{z'}(\underline{W}^{z'}),\underline{W}^{z}) > t) -Pr(T(z',\underline{M}^z(\underline{W}^z),\underline{W}^{z}) > t)$$ or modify the definition of $\NIE_W(t,z,z')$ as  
%$$\NIE_W(t,z,z')=Pr(T(z',\underline{M}^{z}(\underline{W}^{z}),\underline{W}^{z'}) > t) -Pr(T(z',\underline{M}^z(\underline{W}^z),\underline{W}^{z}) > t)$$. I think modify the definition of $\NIE_W(t,z,z')$ might be more reasonable.]
%}

%{\color{blue}[Please add word explanations for these effect. Specifically, which pathways are contained in each of these quantities. The current color in figure 1 seems not related to pathway and thus does not explain the effects as in my previous comments below: ]]}. 
%{\color{blue}[NIE is not defined. Need more discussion on the meaning of these different indirect effects. $NIE_W$ here seems to be the direct effect through $W$ rather than all mediation effect through $W$.I think we will need a graph using different color to quantify different pathways included in each of the effects.}

Similarly, for $\NDE$, $\NIE_M$ and $\NIE_W$ among subgroups with covariates $X=x$, we can define them as
\begin{align*}
\NDE(t,z,z',z'',x)&=Pr(T(z',\underline{M}(z,\underline{W}(z)),\underline{W}(z'')) > t|X=x) \\
&-Pr(T(z,\underline{M}(z,\underline{W}(z)),\underline{W}(z'')) > t|X=x),\\
\NIE_M(t,z,z',z'',x)&=Pr(T(z',\underline{M}(z',\underline{W}^{z'}),\underline{W}(z'')) > t|X=x) \\
&-Pr(T(z',\underline{M}(z,\underline{W}(z)),\underline{W}(z'')) > t|X=x),\\
\NIE_W(t,z,z',z'',x)&=Pr(T(z',\underline{M}(z'',\underline{W}(z'')),\underline{W}(z')) > t|X=x) \\
&-Pr(T(z',\underline{M}(z'',\underline{W}(z'')),\underline{W}(z)) > t|X=x).
\end{align*}
%{\color{blue}[This can also be modified to be the same as previous sets of equation so that the TE will be a summation of these 3 effects.]}

\subsection{General models and determination of causal mechanism}
To estimate the effects, we assume the following joint models for this casual mechanism:

\begin{equation}
\label{eq:equ_M}
r_i(z,\underline{w},t)=r_{0}(t)\exp\{\beta_z z+\bmbeta_x^\top\bmX_i+\bmbeta_w w(t)+\nu_i+\gamma u_i\},
\end{equation}
\begin{equation}
\label{eq:equ_W}
W_i(z,t)=\alpha_0+\alpha_zz+\bmalpha_x^\top\bmX_i+\alpha_tt+u_i+\varepsilon_i(t),
\end{equation}
\begin{equation}
\lambda_i(z,\underline{m},\underline{w},t)=\lambda_0(t)\exp\{\eta_z z+\eta_mm(t)+\eta_ww(t)+\bmeta_x^\top\bmX_i +\delta_1\nu_i+\delta_2 u_i\},
\label{eq:surv}
\end{equation}
where $\nu_i\sim_{i.i.d.} N(0,\sigma_\nu^2)$ are shared random effects between the mediator $M$ and the outcome that are independent of $\bmX$, $Z$, and $u_i\sim_{i.i.d.} N(0,\sigma_u^2)$ are shared random effects among mediators $M$, $W$ and the outcome that are independent of $\bmX$, $Z$. Here $r_i(z,\underline{w},t)$ is the intensity function for the recurrent event process $M_i(z, \underline{w}, t)$. Mean for the process $W_i(z,t)$ is linearly associated with the treatment indicator $z$ with coefficient $\alpha_z$, some other known covariates $\bmX_i$ with coefficient $\alpha_x$, the time $t$ with coefficient $\alpha_t$ and intercept $\alpha_{0}$. The residual term $\varepsilon_i(t)$ is the individual specific biomarker fluctuation over time plus measurement error (if any), which is assumed to be independent of $z$, $\bmX_i$, but can be dependent on $t$. $\lambda_i(z,\underline{m},\underline{w},t)$ is the hazard function for the potential time to terminal event $T(z,\underline{m},\underline{w})$. Here we make a Markov assumption in equations \ref{eq:equ_M} and \ref{eq:surv} so that the intensity function and hazard will only depend on current $w(t)$ and $m(t)$. However, this assumption is not crucial and the method can be extended straightforwardly to allow a finite number of pre-specified lag time effects in the right-hand side of these two equations.

\subsection{Relaxation of Sequential Ignorability}
Sequential Ignorability (SI) is a fundamental assumption in mediation analysis,  ensuring that there is no unmeasured confounding among the treatment, mediators, and outcome variables. The traditional SI assumption requires that no confounding between the exposure and the mediator process after conditioning on observed covariates, as well as no confounding between the mediator process and the outcome after conditioning on observed covariates and the exposure. In cases where unmeasured confounding exists between the mediators and the outcome of interest, a weaker version of the SI assumption can be applied. This relaxed assumption requires that the independence between the mediators and the outcome holds only after conditioning on observed covariates, treatment, and all shared random effects with a specified distribution \citep{Niu2023}. This modification allows the use of shared random effects to account for unmeasured time-independent confounders. In our study, we extend the SI assumption to accommodate multiple mediators, repeated measurement of mediators, and recurrent events. This extension ensures that the assumption remains valid when dealing with multiple causally dependent mediators, i.e., 
\begin{equation} \label{eqn:SI-1}
Z~\indep (\underline{\bmM}(z',\underline{\bmW}^{z'}),\underline{\bmW}^{z''},T(z,\underline{\bmm},\underline{\bmw}))|\bmX \text{ for all  } z,z',z'', \underline{\bmm},\underline{\bmw},
\end{equation}
\begin{equation} \label{eqn:SI-2}
\underline{\bmW}^{z''}~\indep (\underline{\bmM}(z',\underline{\bmW}^{z'}),T(z,\underline{\bmm},\underline{\bmw}))|Z,\bmX, u \text{ for all  }  z,z',z'',\underline{\bmm}, \underline{\bmw},
\end{equation}
\begin{equation} \label{eqn:SI-3}
\underline{\bmM}^{z'}~\indep T(z,\underline{\bmm},\underline{\bmw})|Z,\bmX,u,\nu, \text{ for all}  z,z',z'',\underline{\bmm},\underline{\bmw}.
\end{equation}

The first part of the sequential ignorability requires that $Z$ is ignorable given the measured covariates $X$. This assumption is automatically guaranteed when the exposure of interest is the treatment under a randomized trial. 
For the second part, the alternative mediator $W$ is ignorable after conditioning on treatment $Z$, covariates $X$, and the shared random effect $u$ among mediators and outcomes. 
For the third part, the mediator of interest M is assumed to be independent after conditioning on the treatment Z, covariates X and all shared random effects $u$ and $\nu$.

Under this relaxed assumption, we can write the survival function for $T_i(z,\underline{M}(z',\underline{W}^{z'}),\underline{W}^{z''})$  %{\color{red}[I think we might also provide a formula here or in appendix for $S(t,z,z',z'';x)$ to identify all pathway effects, which provides the survival function for $T_i^{z\underline{M}_i^{z'}(W_i^{z'})\underline{W}_i^{z'''}}$]}
\begin{eqnarray*}
&&{\color{black}S(t,z,z',z'';\bmx)}\\
&=&{\color{black}\int_{\nu}\int_{u}\int_{\underline{w}}\int_{\underline{w'}} \int_{\underline{m}}S(t|Z=z,\underline{M}=\underline{m},\underline{W}=\underline{w},\bmX=\bmx,\nu,u)dF_{\underline{M}|Z,\bmX,\bmW,\nu,u}(\underline{m}|z',\bmx,\bmw',\nu,u)}\\
&&{\color{black}dF_{\underline{W}|Z,\bmX,u}(\underline{w}|z'',\bmx,u)dF_{\underline{W}|Z,\bmX,u}({\color{black}\underline{w'}}|z',\bmx,u)dF_{u}dF_{\nu},}
\end{eqnarray*}
%\begin{eqnarray*}
%S(t,z,z';\bmx)&=&\int_{\nu}\int_{u}\int_{\underline{w}} \int_{\underline{m}}S(t|Z=z',\underline{M}=\underline{m},\underline{W}=\underline{w},\bmX=\bmx,\nu,u)dF_{\underline{M}|Z,\bmX,\bmW,\nu,u}(\underline{m}|z,\bmx,\bmm,\nu,u)dF_{\underline{W}|Z,\bmX,u}(\underline{w}|z,\bmx,u)dF_{u}dF_{\nu},
%\end{eqnarray*}
%{\color{red}[Need give the interpretation of this quantity, seems to be survival function for $T_i^{z'\underline{M}^z(\underline{W}^z)\underline{W}^{z'}}$]}
where $S(t|Z=z,\underline{M}=\underline{m},\underline{W}=\underline{w},\bmX=\bmx,\nu,u)$ is the conditional survival probability given the exposure Z, the mediators W and M, the covariates X, and the shared random effects $\nu$ and $u$; $dF_{\underline{M}|Z,\bmX,\bmW,\nu,u}(\underline{m}|z',\bmx,\bmw',\nu,u)$ is the probability density function for the whole mediator process $\underline{M}$ given the exposure, the covariates, the alternative mediator $w$ and the shared random effects $\nu$ and $u$. $dF_{\underline{W}|Z,\bmX,u}({\color{black}\underline{w}}|z'',\bmx,u)$ is the probability density function for the whole mediator process $\underline{W}$ given the exposure, the covariates, and the shared random effect $u$. All terms are no longer dependent on potential outcomes and thus are estimable from the data. 

Specifically, considering the model with the shared random effects, we can write 
\begin{eqnarray*}
&&S(t|Z=z,\underline{M}=\underline{m},\underline{W}=\underline{w},\bmX=\bmx, \nu, u)\\
&=&\exp\left\{-\exp{(\eta_z z+\bmeta_x^\top\bmX+\delta_1 \nu_i +\delta_2 u_i)}\int_{0}^{t} \lambda_0(u)\exp{(\eta_m m(u) + \eta_w w(u))}\,du\right\}.
\end{eqnarray*}
and 
\begin{eqnarray*}
\label{eq:df}
dF_{\underline{M}|Z,\bmX,\bmW,\nu,u}(\underline{m}|z',\bmx,\bmw',\nu,u)&=&\prod_{j=1}^{J(\underline{m})}[r_{0}(R_{j}(\underline{m}))\exp(\beta_z z'+\bmbeta_x^\top\bmx+\beta_w w'(R_{j}(\underline{m}))+\nu+\gamma u) \nonumber]\\
&&\times\exp\left\{-\int_{0}^{\tau} r_0(s)\exp(\beta_z z'+\bmbeta_x^\top\bmx+\beta_w w'(s) + \nu+\gamma u)ds\right\},\\
dF_{\underline{W}|Z,\bmX,u}(\underline{w}|z'',\bmx,u)&=&\prod_{j=1}^{J(\underline{w})}[\frac{1}{\sigma_\varepsilon\sqrt{2\pi}}\exp\{\frac{-\varepsilon_j^2}{2{\sigma_{\varepsilon}}^2}\}d\nu],\\
dF_{\nu}(\nu)&=&\frac{1}{\sigma_\nu\sqrt{2\pi}}\exp\{\frac{-\nu^2}{2{\sigma_{\nu}}^2}\}d\nu,\\
dF_{u}(u)&=&\frac{1}{\sigma_u\sqrt{2\pi}}\exp\{\frac{-u^2}{2{\sigma_{u}}^2}\}du,
\end{eqnarray*}

where $J(\underline{m})$ is the number of jumping points for recurrent events within $[0,\tau]$; 

$R_{1}(\underline{m}), \cdots, R_{J(\underline{m})}(\underline{m})$ are corresponding jumping time points; $\varepsilon_j=w_j-\alpha_0-\alpha_z z'' -\bmalpha_x^\top\bmx-u$; 
$J(\underline{w})$ is the number of jumping points for w within $[0,\tau]$; 
 $F_{\nu}$ is the cumulative density function of $\nu$ from the normal distribution with mean 0 and variance $\sigma_{\nu}^2$; $F_{u}$ is the cumulative density function of $u$ from the normal distribution with mean 0 and variance $\sigma_{u}^2$.

To get the estimated version of $\hat{S}(t,z,z',z'';\bmx)$, we can obtain estimators $\hat{S}(t|Z=z,\underline{M}=\underline{m},\underline{W}=\underline{w},\bmX=\bmx)$, $d\hat{F}_{\underline{M}|Z,\bmX,\bmW,\nu,u}(\underline{m}|z',\bmx,\bmw',\nu,u)$, $d\hat{F}_{\underline{W}|Z,\bmX,u}(\underline{w}|z'',\bmx,u)$ by plugging in the estimates.

\subsection{Numerical estimate for $\NDE$, $\NIE_W$ and $\NIE_M$}
We rely on the survival function introduced in section 2.2 to compute $\NDE$, $\NIE_W$, and $\NIE_M$. However, since $\underline{\bmm}$ and $\underline{\bmw}$ have infinite dimensions, the required integrals for computing these effects do not have a closed-form solution. This complexity arises from the infinite-dimensional nature of the variables, making traditional analytical integration methods infeasible. To approximate the integral, we use the Monte Carlo method, which relies on random sampling to estimate the result. 

We first use maximum likelihood estimation to estimate the parameters $r_0(t),\lambda_0(t),\bmbeta_x^\top,\beta_z,\beta_w$,
$\gamma,\alpha_0, \alpha_z, \bmalpha_x^\top,\alpha_t, \bmeta_x^\top,\eta_z,\eta_m,\eta_w,\delta_1,\delta_2$. We use Gaussian
quadrature tools to maximize the joint likelihood. This method is available in standard statistical packages such as Proc NLMIXED in SAS. All the details regarding the joint likelihood and underlying assumptions are specified in Appendix Section A. 

We next sample random effects, including $\nu\sim N(0,\sigma_\nu^2)$, $u\sim N(0,\sigma_u^2)$ and $\varepsilon (t_{j(\underline{w})}(\underline{w})) \sim N(0,\sigma_\varepsilon^2) $ at each time $t_{j(\underline{w})}(\underline{w})$. $t_{j(\underline{w})}(\underline{w})$ is the corresponding $j$th jumping time point for repeated measures within $[0,\tau]$. Then we generate repeated measures and recurrent events mediators by first sampling the repeated measures mediators $\bmW^1(t_{j(\underline{w})}(\underline{w}))$ and $\bmW^0(t_{j(\underline{w})}(\underline{w}))$ for $t_1(\underline{w}), \cdots, t_{J(\underline{w})}(\underline{w})$ and the recurrence event process $\bmM^1(s,\underline{W}^1 (s))$ and $\bmM^0(s,\underline{W}^0 (s))$ for $s\in [0,t]$ using the inverse approach \citep{cinlar1975}. Finally, we can estimate the 8 survival probabilities $\hat{S}(t|z,\underline{\bmM}^
{z'}(\underline{\bmW}^{z'}),\underline{\bmW}^{z''},\bmX,\nu,u)$ for $z, z', z''\in \{0,1\}$.

Alternatively, potential survival times $T(z,\underline{\bmM}(z',\underline{\bmW}^{z'}),\underline{\bmW}^{z''})$ can be directly sampled. This alternative approach is computationally useful, especially when we are interested in the survival functions at multiple time points.
In the current study, we compute all the combinations of $\hat S(\cdot)$ via numerical integration and then calculate the $\NDE$, $\NIE_W$, and $\NIE_M$ based on these survival probabilities. The detailed method is stated in Web Appendix B.  
Confidence intervals are computed using Bootstrap. To evaluate the robustness of the Monte Carlo integration method, we conducted the results by varying the number of nodes used for the integration and the number of time points. This helps to evaluate the Monte Carlo error and ensures the accuracy of our estimates. 

\section{Simulation}
\label{s:sim}
To evaluate the performance of our models' parameter estimation and the performance/robustness of the effects estimators $\NDE$, $\NIE_W$, $\NIE_M$, we conducted simulations with two settings.
In Setting I, 200 datasets were simulated according to our proposed models. The details on the simulation setting can be found in Web Appendix C.
Using the numerical estimation method stated in Section 2.3, we fit our model for each simulated dataset and apply the parameter estimates to $\NDE$, $\NIE_W$, and $\NIE_M$ calculations. 
Table \ref{tab:1} shows the simulation results for $\NDE$, $\NIE_W$,$\NIE_M$. When the random effects follow normal distribution $\nu \sim N(0,0.49)$, $ u \sim N(0,1)$ and  $\epsilon(t)\sim N(0,1)$ for $t=(0,2,4,6,8)$, the joint model is correctly specified, the fitting for both $\NDE$, $\NIE_M$ perform well with small bias and approximate correct coverage rate. 
%However, a slightly lower coverage is observed for $\NIE_W$ at early time points, which is likely due to the small true value. In this case, even if the absolute bias is small, the relative bias can be large, which contributes to the lower coverage.

We also evaluate the robustness of our proposed estimators to the misspecification of the model. We misspecified the random effect $\nu$ using log gamma distribution with shape parameter 2 and scale parameter 1.8, and misspecified the random effect $u$ using log gamma distribution with shape parameter 1.5 and scale parameter 1 (settings II). From Table \ref{tab:1}, we can see our estimator is sensitive to this type of misspecification, especially for later time points. The results remind us that our proposed estimator is sensitive to the misspecification of random effect distribution. This suggests the necessity of evaluating the goodness of fit in real data analysis. 

%{\color{blue}[We can keep this as is for thesis. However, this might be tricky part for journal submission. It might be better to reduce the variance and show some kind of robustness. Otherwise, we might need to perform the goodness of fit test in real data analysis. I don't know how the random effect distribution are usually tested, maybe we can ask Dr. Liu.]}

\section{Real data analysis}
\label{s:analysis}

\subsection{CPCRA Study}\label{ss:CPCRA}

The CPCRA study was originally designed to evaluate the relative efficacy and safety of didanosine (ddI) and zalcitabine (ddC) in HIV-infected patients who could not tolerate zidovudine (AZT). In this clinical trial, 467 patients who had previously received AZT and had CD4 cell counts of 300 or fewer per cubic millimeter were recruited and randomly assigned to receive either ddI (n=230) or ddC (n=237). At baseline, the following variables were assessed: (1) previous AIDS-defining condition (PADC; 1: AIDS diagnosis at baseline; 0: no AIDS diagnosis); (2) gender (1: female; 0: male); (3) treatment (1: ddI; 0: ddC); (4) stratum of response to AZT (1: AZT intolerance; 0: AZT failure); (5) baseline hemoglobin (centered at mean = 12). CD4 cell counts are commonly used as a biomarker of disease status in HIV infection. Over the 21-month follow-up period, CD4 counts were measured repeatedly at baseline and every two months. However, due to loss to follow-up or missed visits, CD4 count data exhibited varying levels of missingness. Especially for months 4, 8, and 10, there are more than 65 $\%$ data missing. Each participant had between 1 and 8 CD4 measurements, with a median of 4. The high level of data missing for repeated CD4 counts may affect the accuracy of our estimation. However, our model allows each individual to have its own time jump points to handle this situation %{\color{red} why not do a last value carried forward imputation?     }. 
The primary outcome of interest in this study was overall survival. A total of 100 and 88 deaths occurred in the ddI and ddC groups, respectively.  We aimed to investigate the mediating role of opportunistic infection (OI) recurrence and repeated CD4 count measurements in the effect of treatment on survival using our models. The diagnosis times of OIs were recorded and served as the recurrent events mediator ($M$), while repeated CD4 count measures, including baseline values, were considered as the repeated measure mediator ($W$) in our model. Two variables were separately considered as exposure in this Section. In subsection 4.2, the treatment (ddI/ddC) was the primary exposure of interest, while the other four variables were covariates in recurrent events, repeated CD4 measurement, and survival models. In subsection 4.3, PADC was the primary exposure of interest, while treatment was included as one of the covariates.

Before applying our models, we first verified whether the required assumptions were met. Regarding the first part of the SUTVA assumption, the exposure variables (treatment and PADC) were clearly defined without subcategories, and no interference between individuals was expected.  That is, an individual's treatment assignment, PADC status, OI development, or repeated CD4 count measurements were unlikely to influence another individual's outcome. We also assumed that data were collected accurately and in a timely manner, meeting the consistency assumption. 
For the first part of the SI assumption, randomization in the clinical trial ensured its validity for treatment assignment. For PADC, we considered the assumption reasonable after adjusting for all covariates. For parts II and III, we allowed some relaxation, assuming that after modeling unmeasured time-independent confounders using shared random effects, the ignorability of mediators (repeated CD4 counts and recurrent OIs) would hold given observed covariates, exposure, and quantified unmeasured confounding.

\subsection{Comparative effectiveness between the two treatment arms}\label{CPCRA}
Didanosine (ddI) and Zalcitabine (ddC) were second-line of HIV treatments used for patients who could not tolerate AZT or experienced worsening clinical conditions during AZT treatment \cite{Abrams1994}. In this section, we focus on understanding the mediation roles of two mediators, recurrent OIs and repeated CD4 counts in the pathway linking ddI and ddC treatment to patient survival. To analyze these relationships, we fitted the data using joint models incorporating a recurrent events model \eqref{eq:equ_M}, a repeated measurements model \eqref{eq:equ_W}, and a survival model \eqref{eq:surv}. The baseline hazard function was modeled as a piecewise constant function with four intervals. The jump points are determined by the 25th, 50th, and 75th quartiles among all patients' terminal times (death or exit the study). Similar to the terminal events, the baseline intensity for recurrent events was also modeled as a piecewise constant function with jump points determined by 25th, 50th, and 75th quartiles of all recurrent events time. The estimated parameters are presented in Table \ref{tab:2}. Notably, the estimated baseline hazard rate for OI recurrence decreased over time, whereas the baseline hazard rate for survival events increased.
After adjusting for PADC, gender, stratum of response to AZT, and baseline HgB, we found that ddC had a similar effect to ddI on repeated CD4 counts ($p=0.24$), recurrent events ($\text{Log Hazard Ratio (LHR)} = 0.01$, $p=0.92$), and survival ($\text{LHR} = -0.23$, $p=0.28$). To further examine the direct and indirect effects of treatment on survival, we estimated the $\NDE$, $\NIE_{OI}$, $\NIE_{CD4}$ and $TE$ using the joint modeling. The estimated values (solid line) along with the bootstrapped 95\% pointwise confidence intervals (dashed lines) are presented in Figure \ref{fig:2} A and B. 
Our findings indicate that neither the $\NDE$, $\NIE_{OI}$, nor $\NIE_{CD4}$ were significantly different from zero, suggesting that treatment (ddC vs. ddI) had no significant direct or indirect effect on survival. These results align with our parameter estimates but differ from previous findings. In a prior study \cite{Niu2023}, treatment was reported to have a significant direct effect on survival. This discrepancy likely arises from differences in model specification. Unlike the previous study, which only considered recurrent events as a mediator, our study additionally incorporated repeated CD4 count measurements as a second mediator. So the previous study’s definition of NDE implicitly included both NDE and $\NIE_W$ from our model, meaning their reported significant direct effects were actually the mediation effect through CD4 counts. By explicitly separating these effects, our model provides a more flexible and accurate evaluation of treatment effects.
Additionally, we performed a Cox regression analysis to estimate the total effect of treatment, represented by the green line in Figure \ref{fig:2}A. The TE estimated from the Cox model closely matched the sum of NDE, $\NIE_{OI}$, and $\NIE_{CD4}$, demonstrating strong consistency between our joint modeling approach and traditional Cox regression. 

%There may be servaral reasons that we didn't see significant effects of ddC on survival outcome. First, we included repeated CD4 counts as another mediator besides recurrent events in the survival model. Both number of OIs and repeated CD4 counts show significant effects on survival outcome. After adjusting these two mediators, the effect of ddC may get less significant.

%Moreover, tt can be seen that the random effects $u$ is showing the significance in survival model ($p<0.001$), which indicates there are unmeasured confounding between repeated CD4 counts and survival models and could affect the significance of treatment too. 

 \subsection{Effect of prior AIDS-defining conditions}\label{CPCRACD4}
AIDS-defining conditions include opportunistic infections and cancers that become life-threatening in individuals with HIV \citep{AIDS_C}. The CDC defines these conditions to classify the stages of HIV progression and monitor disease advancement \citep{AIDS}. Once a patient experiences their first AIDS-defining event, they are diagnosed with AIDS and enter Stage 3 of HIV infection. In the CPCRA study, the variable PADC reflects whether a patient had a previous AIDS-defining condition. Our analysis aimed to evaluate the impact of PADC on survival and its mediation effects through recurrent opportunistic infections (OIs) and repeated CD4 count measurements. The results indicate that PADC had a significant effect on survival ($\text{LHR}=1.80$, $p<0.001$), repeated CD4 counts (Est=$-0.60$, $p<0.001$), and recurrent events ($\text{LHR}=0.45$, $p=0.03$). These findings highlight the importance of PADC as an indicator of HIV disease progression, with patients having a prior AIDS-defining condition exhibiting lower repeated CD4 counts, a higher risk of recurrent OIs, and an increased risk of mortality.

To better understand these relationships, we examined the role of baseline covariates. Lower hemoglobin (HgB) levels were significantly associated with lower repeated CD4 counts ($p<0.001$), higher risk of recurring OIs ($p=0.002$), and higher mortality ($p<0.001$), consistent with previous findings \citep{Suja2020, Wang2021}. Gender differences were also observed, with male patients showing lower repeated CD4 counts ($p=0.02$) and a higher probability of developing OIs ($\text{LHR}=-0.69$, $p=0.01$), though survival time was similar between males and females ($\text{LHR}=-0.35$, $p=0.37$). 

For the repeated measurements model, CD4 counts declined over time ($p<0.001$), indicating disease progression. In the survival model, the estimated log hazard ratio for repeated CD4 counts as a mediator $W(t)$ was $0.77$ ($p=0.02$), suggesting a positive association between repeated CD4 counts and mortality. This may differ from previous studies since lower CD4 counts are typically associated with higher mortality. However, CD4 counts were also included as a covariate in the recurrent events model, which may partially account for their indirect effect through OIs. The number of OIs as a mediator $M(t)$ was strongly associated with mortality ($\text{LHR}=0.43$, $p<0.001$), reinforcing the critical role of recurrent OIs in HIV progression and survival. Furthermore, the shared latent effects of $u$ were significant in the survival model, indicating that the effect of repeated CD4 counts on survival was confounded by unmeasured shared factors.

\begin{figure}
\includegraphics[scale=0.9]{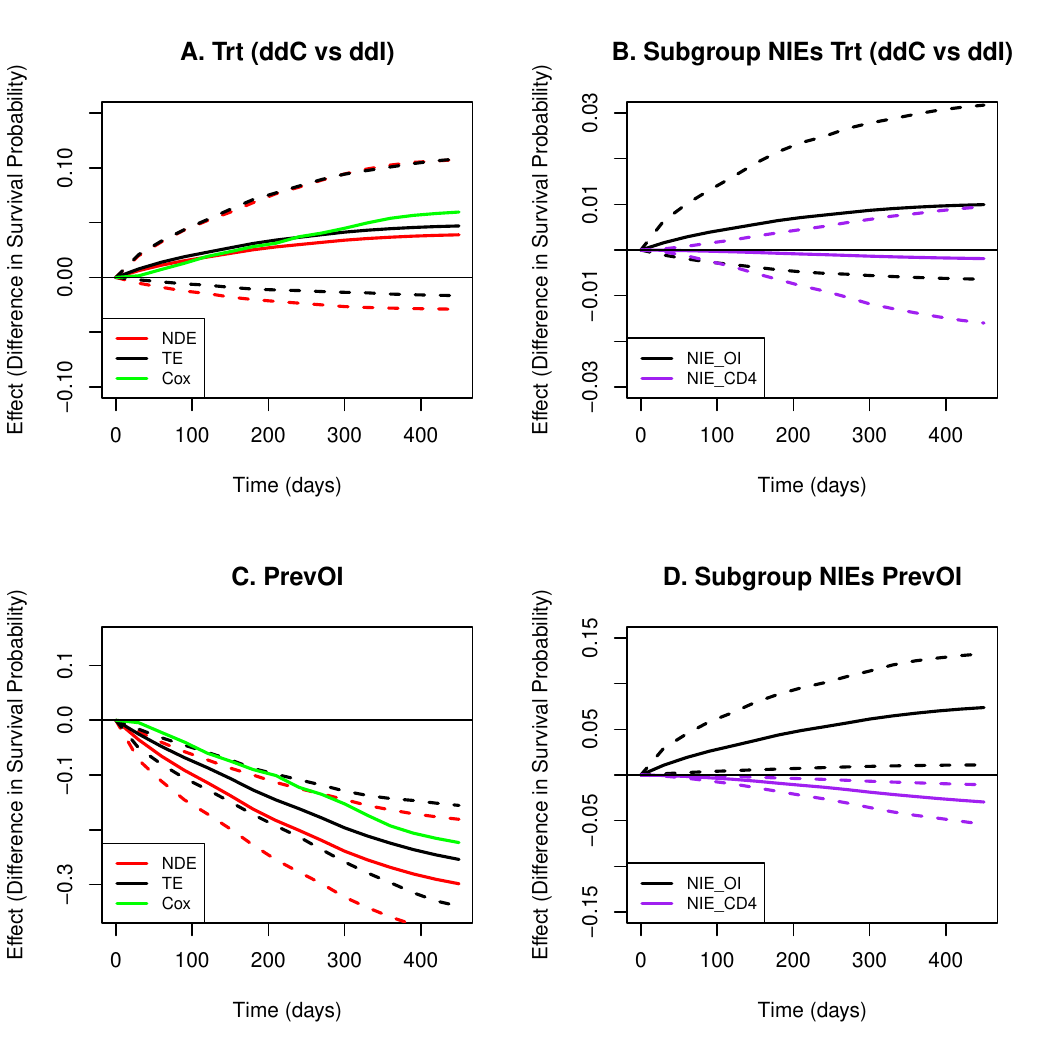}
\caption{Estimation (solid lines) with bootstrapped 95\% point-wise confidence intervals (dash lines) of $\NDE$, $\NIE_{OI}$, $\NIE_{CD4}$ and TE of Treatment (ddC vs ddI) and PADC (prior AIDS-defining conditions) on overall survival probability for CPCRA study using our model and alternative methods.}
\label{fig:2}
\end{figure}

To further explore the direct and indirect effects of PADC on survival, we estimated the natural direct effect ($\NDE$), the natural indirect effect through OIs ($\NIE_{OI}$), and the natural indirect effect through repeated CD4 counts ($\NIE_{CD4}$).
The estimated effects with bootstrapped 95\% confidence intervals (dashed lines) are shown in Figure \ref{fig:2} (C and D). The $\NDE$ was significantly different from zero over the follow-up period, suggesting that PADC had a strong direct effect on survival. Both $\NIE_{OI}$ and $\NIE_{CD4}$ were significant over time but had opposite directions, indicating that both OIs and repeated CD4 counts mediated the impact of PADC on survival. However, the magnitude of $\NIE_{OI}$ was larger than $\NIE_{CD4}$, suggesting that opportunistic infections played a more dominant role in mediating the effect of PADC on survival.

Finally, we {\color{black} confirmed the consistency of our model with the Cox regression model in terms of the estimated total effect (TE) of PADC on survival probability}, shown as the green line in Figure \ref{fig:2}. The TE estimated from the Cox model closely matched the sum of $\NDE$, $\NIE_{OI}$, and $\NIE_{CD4}$, indicating strong consistency between our joint modeling approach and traditional Cox regression.

\begin{table}[hbt!]
 \centering
  \caption{True Values, Bias, empirical standard deviation (SD), median estimated standard error (MeSE), and coverage rate for 95\% nominal confidence interval (CR) from two simulation settings.}
\begin{tabular}{cccccccc}
\hline
Setting &Time &Effect &True Value &Bias &SD &MeSE &CR\\
\hline
I &2 &NDE &-0.036 &0.001 &0.026 &0.026 &96.0\%\\
 & &$\NIE_M$ 
&-0.013 &0.000 &0.008 &0.008 &94.0\%\\
& &$\NIE_W$ 
&-0.007 &0.000 &0.004 &0.004 &98.0\%\\
 &4 &NDE &-0.081 &0.001 &0.056 &0.056 &94.0\%\\
  & &$\NIE_M$ 
&-0.028 &-0.001 &0.018 &0.017 &94.0\%\\
& &$\NIE_W$ 
&-0.037 &-0.003 &0.019 &0.020 &97.0\%\\
 &6 &NDE &-0.078 &0.001 &0.054 &0.053 &94.0\%\\
  & &$\NIE_M$ 
&-0.027 &0.000 &0.017 &0.017 &94.5\%\\
& &$\NIE_W$ 
&-0.044 &-0.005 &0.023 &0.024 &96.5\%\\
 &8 &NDE &-0.054 &0.002 &0.038 &0.038 &94.0\%\\
  & &$\NIE_M$ 
&-0.019 &0.000 &0.012 &0.012 &93.0\%\\
& &$\NIE_W$ 
&-0.031 &-0.004 &0.017 &0.018 &96.5\%\\
\hline
II &2 &NDE &-0.036 &-0.030 &0.042 &0.048 &83.4\%\\
 & &$\NIE_M$ 
&-0.013 &-0.012 &0.018 &0.015 &86.6\%\\
& &$\NIE_W$ 
&-0.007 &-0.028 &0.015 &0.012 &38.0\%\\
 &4 &NDE &-0.081 &0.021 &0.038 &0.044 &87.2\%\\
  & &$\NIE_M$ 
&-0.028 &0.006 &0.015 &0.013 &81.3\%\\
& &$\NIE_W$ 
&-0.037 &-0.011 &0.021 &0.017 &82.4\%\\
 &6 &NDE &-0.078 &0.049 &0.019 &0.020 &35.3\%\\
  & &$\NIE_M$ 
&-0.027 &0.016 &0.007 &0.006 &33.2\%\\
& &$\NIE_W$ 
&-0.044 &0.019 &0.011 &0.009 &47.1\%\\
 &8 &NDE &-0.054 &0.041 &0.010 &0.009 &9.6\%\\
  & &$\NIE_M$ 
&-0.019 &0.014 &0.004 &0.003 &14.4\%\\
& &$\NIE_W$ 
&-0.031 &0.019 &0.007 &0.004 &16.6\%\\
\hline

\end{tabular}
\label{tab:1}
\end{table}

\begin{table}
 \centering
  \caption{Fitted regression parameters from the data analysis of CPCRA study.}
\begin{tabular}{cccc}
\hline
\
&\multicolumn{3}{c}{Results} \\
\hline
Variable & Est & SE &p-value\\
\hline
&\multicolumn{3}{c}{Repeated measuring of CD4 counts}\\
\hline
Intercept $\alpha_0$ &1.12&0.11&$<$0.001\\
Treatment &-0.08&0.07&0.24\\
PADC &-0.60&0.09&$<$0.001\\
Gender &0.27&0.12&0.020\\
Stratum of Response to AZT &0.03&0.08&0.71\\
HgB &0.16&0.02&$<$0.001\\
Time &-0.03&0.002&$<$0.001\\
$\log(\sigma^2_{\epsilon})$ &-1.72&0.04&$<$0.001\\
\hline
&\multicolumn{3}{c}{Recurrent Event}\\
\hline
Treatment &0.01&0.11&0.92\\
PADC &0.45&0.21&0.03\\
Gender &-0.69&0.26&0.01\\
Stratum of Response to AZT &-0.08&0.13&0.53\\
HgB &-0.16&0.05&0.002\\
Repeated CD4 &-0.29&0.23&0.21\\
Shared random effect ($u$) $\gamma$ &-0.49&0.28&0.08\\
$\log(\sigma^2_{\nu})$ &-3.06&1.84&0.10\\
$\log(\sigma^2_u)$ &-0.82&0.07&$<$0.001\\
$\log(r_0(t))$: $t\in[0.03,2.77)$ &-2.83&0.33&$<$0.001\\
$\log(r_0(t))$: $t\in[2.77,6.33)$ &-2.95&0.32&$<$0.001\\
$\log(r_0(t))$: $t\in[6.33,10.20)$ &-2.88&0.30&$<$0.001\\
$\log(r_0(t))$: $t\in[10.20,20.10)$ &-2.96&0.28&$<$0.001\\
\hline
&\multicolumn{3}{c}{Survival}\\
\hline
Treatment &-0.23&0.21&0.28\\
PADC &1.80&0.40&$<$0.001\\
Gender &-0.35&0.38&0.37\\
Stratum of Response to AZT &-0.10&0.25&0.69\\
HgB &-0.58&0.10&$<$0.001\\
Number of OIs &0.43&0.11&$<$0.001\\
Repeated CD4 &0.77&0.33&0.02\\
Shared Random Effect ($\nu$) 
&0.38&1.54&0.80\\
Shared Random Effect ($u$) &-2.35&0.50&$<$0.001\\
$log(\lambda_0(t))$: $t\in[0.47,4.58)$ &-6.19&0.65&$<$0.001\\
$\log(\lambda_0(t))$: $t\in[4.58,8.60)$  &-5.77&0.58&$<$0.001\\
$\log(\lambda_0(t))$: $t\in[8.60,11.62)$ &-5.31&0.56&$<$0.001\\
$\log(\lambda_0(t))$: $t\in[11.62,19.07)$ &-5.20&0.54&$<$0.001\\

\hline
\end{tabular}
\vspace{0.3in}
\label{tab:2}
\end{table}

\section{Discussion}
\label{s:discuss}

In this study, we extended the previous joint modeling framework by incorporating multiple types of mediators - the recurrent events mediator and the repeated measures mediator, and analyzed their mediation effects in the causal pathway from the  exposure to a survival outcome. Previous studies have typically used joint modeling approaches to examine either recurrent events and survival outcomes or repeated measurements and survival outcomes separately \citep{Niu2023, zheng2021}. However, in real analyses, it is possible for both types of mediators to coexist and play crucial roles in the causal pathway. Our current study integrates both types of mediators within a joint modeling framework, providing flexibility to examine the exposure, two mediators, and the survival outcome simultaneously. This novel extension enhances our ability to assess complex mediation effects in survival analysis.

We applied our proposed method to the CPCRA study, considering treatment and previous AIDS-defining conditions (PADC) as the exposure variables separately. We estimated the mediation effects of recurrent opportunistic infections (OIs) and repeated CD4 count measures, both of which play critical roles in HIV progression. Our findings suggest that these two mediators influence the effect of PADC on patient mortality in opposite directions. This insight provides strategic guidance for managing patient health outcomes—particularly for individuals who have already developed AIDS. Specifically, controlling recurrent OIs is more important in the regards of improving patients' survival.

Interestingly, compared to the previous chapter, we did not observe a significant direct effect of treatment on survival outcomes. This highlights an advantage of our approach. In prior studies, the Natural Direct Effect (NDE) implicitly included indirect effects mediated through CD4 counts, making it challenging to isolate the direct impact of exposure. By explicitly modeling CD4 as an additional mediator, our framework distinguishes between the direct {\color{black} effect of exposure on survival probability and those indirect effects mediated through CD4 counts}, offering a more detailed understanding of their respective contributions. Given that CD4 levels influence the likelihood of recurrent infections, our model also captures the dynamic interplay between these two mediators.

Moreover, our framework relaxes the Sequential Ignorability (SI) assumption by incorporating shared latent random effects to account for unmeasured confounders. This is particularly important because PADC are not randomized, and their relationship with CD4 counts and survival may be influenced by unmeasured factors such as immune response variability, treatment adherence, or other comorbidities. Future research could explore alternative causal inference techniques, such as instrumental variables or sensitivity analyses, to better account for unmeasured confounding.

Our study presents a joint modeling approach capable of handling multiple types of mediators, allowing the mediator $\bmM$ to be causally affected by the alternative mediator $\bmW$. However, in the real-world setting, mediators may influence each other bidirectionally. Addressing this complexity may require the integration of causal structural models, such as mutual causal models or structural equation modeling (SEM)\citep{Grace2021, Gunzler2013}, to better capture the interdependent relationships between mediators. Future works focusing on developing methods to account for such reciprocal causation while maintaining robustness in mediation analysis are needed.

\section*{Acknowledgements}

This research is partly supported by National Institute of General Medical Sciences under grant U54 GM115458, National Heart, Lung, and Blood Institute under grant R01 HL136942, National Institute on Aging grant R21 AG063370 and R01 AG081244, and Washington University Institute of Clinical and Translational Sciences grant UL1TR002345 from the National Center for Advancing Translational Sciences (NCATS).

\section*{Data Availability Statement}
The data that support the findings of this study are from a trial (ClinicalTrials.gov Identifier: NCT00001022) conducted by the CPCRA, which is funded by the National Institute of Allergy and Infectious Diseases (NIAID). Any data request needs to be submitted to the trial PIs (Drs. Saravolatz LD and Winslow DL dwinslow@stanford.edu) and the CPCRA.

\backmatter

%  This section is optional.  Here is where you will want to cite
%  grants, people who helped with the paper, etc.  But keep it short!

%\section*{Acknowledgements}
%  Here, we create the bibliographic entries manually, following the
%  journal style.  If you use this method or use natbib, PLEASE PAY
%  CAREFUL ATTENTION TO THE BIBLIOGRAPHIC STYLE IN A RECENT ISSUE OF
%  THE JOURNAL AND FOLLOW IT!  Failure to follow stylistic conventions
%  just lengthens the time spend copyediting your paper and hence its
%  position in the publication queue should it be accepted.

%  We greatly prefer that you incorporate the references for your
%  article into the body of the article as we have done here
%  (you can use natbib or not as you choose) than use BiBTeX,
%  so that your article is self-contained in one file.
%  If you do use BiBTeX, please use the .bst file that comes with
%  the distribution.  In this case, replace the thebibliography
%  environment below by
%
\bibliographystyle{apa}
\bibliography{bib}

%\begin{thebibliography}{}

%\bibitem{ } Cox, D. R. (1972). Regression models and life tables (with
%discussion).  \textit{Journal of the Royal Statistical Society, Series B}
%\textbf{34,} 187--200.

%\bibitem{ }  Hastie, T., Tibshirani, R., and Friedman, J. (2001). \textit{The
%Elements of Statistical Learning: Data Mining, Inference, and Prediction}.
%New York: Springer.

%\end{thebibliography}

%  If your paper refers to supporting web material, then you MUST
%  include this section!!  See Instructions for Authors at the journal
%  website https://urldefense.com/v3/__http://www.biometrics.tibs.org__;!!JkUDQA!b4l4U2kKGQWWPma5vKzFPaEeungsLE2imnvQw8lOLHYKJ4SYOPupRSatE0Tjggi-XY8$

\section*{Supporting Information}
The Web Appendix referenced in section 2.3 and section 3 is available with this paper on the Biometrics website. Data supporting the findings of this paper can be requested as described in the data availability statement. The SAS and R codes for the simulation and data analysis of this paper are available at \url{https://github.com/nfang-cloud/Joint_modeling_mix_mediator}.

%Put your short appendix here.  Remember, longer appendices are
%possible when presented as Supplementary Web Material.  Please
%review and follow the journal policy for this material, available
%under Instructions for Authors at \texttt{https://urldefense.com/v3/__http://www.biometrics.tibs.org__;!!JkUDQA!b4l4U2kKGQWWPma5vKzFPaEeungsLE2imnvQw8lOLHYKJ4SYOPupRSatE0Tjggi-XY8$ }.

\label{lastpage}

\end{document}